\documentclass[a4paper,fleqn,usenatbib]{mnras}

\usepackage{newtxtext,newtxmath}

\usepackage[T1]{fontenc}
\usepackage{ae,aecompl}

\usepackage{graphicx}	
\usepackage{amsmath}	
\usepackage{amssymb}	

\title[Hydrodynamical Backflow in PKS~2014$-$55]
      {Hydrodynamical Backflow in X-shaped Radio Galaxy PKS~2014-55}

\author[W.~D.~Cotton et al.]{W.~D.~Cotton,$^{1,2}$\thanks{E-mail: bcotton@nrao.edu}
  K.~Thorat,$^{3}$
  J.~J.~Condon,$^{1}$
  B.~S.~Frank,$^{2,4,5}$
  G.~I.~G.~J\'ozsa,$^{2,6,7}$
\newauthor
  S.~V.~White,$^{6}$
  R.~Deane,$^{3}$
  N.~Oozeer,$^{2,8}$
  M.~Atemkeng,$^{9}$
  L.~Bester,$^{2,6}$
  B.~Fanaroff,$^{2}$
\newauthor
  R.~S.~Kupa,$^{2,6}$
  O.~M.~Smirnov,$^{2,6}$
  T.~Mauch,$^{2}$
  V.~Krishnan$^{2}$
  and F.~Camilo$^{2}$
\\
$^{1}$National Radio Astronomy Observatory,
  520 Edgemont Road, Charlottesville, VA 22903, USA\\
$^{2}$South African
  Radio Astronomy Observatory, 2 Fir St, Black River Park,
  Observatory 7925, South Africa\\
$^{3}$Department of Physics, University of Pretoria,
  Private Bag X20, Hatfield 0028, South Africa\\
$^{4}$Department of Astronomy, University of Cape Town, Private Bag
X3,  Rondebosch 7701, South Africa\\ 
$^{5}$Inter-University Institute for Data-Intensive Astronomy,
Department of Astronomy, University of Cape Town, Private Bag X3, \\
~Rondebosch 7701, South Africa\\ 
$^{6}$Department of Physics and Electronics, Rhodes
  University, PO Box 94, Grahamstown 6140, South Africa\\  
$^{7}$Argelander-Institut f\"ur Astronomie, Auf dem
  H\"ugel 71, D-53121 Bonn, Germany\\
$^{8}$African Institute for Mathematical Sciences, 6
  Melrose Road, Muizenberg 7945, South Africa \\
$^{9}$Department of Mathematics, Rhodes
  University, PO Box 94, Grahamstown 6140, South
  Africa\\  
}

\date{Accepted 2020 April 29. Received 2020 April 29; in original form 2020 February 21}

\pubyear{2020}

\begin{document}
\label{firstpage}
\pagerange{\pageref{firstpage}--\pageref{lastpage}}
\maketitle

\begin{abstract}
  We present MeerKAT 1.28 GHz total-intensity, polarization,
{ and spectral-index} images covering the giant (projected length
$l \approx 1.57$~Mpc) X-shaped radio source PKS~2014$-$55 with an
unprecedented combination of brightness sensitivity and angular
resolution. They show the clear ``double boomerang'' morphology of
hydrodynamical backflows from the straight main jets deflected by the
large and oblique hot-gas halo of the host galaxy PGC~064440. The
magnetic field orientation in PKS~2014$-$55 follows the flow lines
from the jets through the secondary wings. The radio source is
embedded in faint ($T_\mathrm{b} \approx 0.5 \mathrm{\,K}$) cocoons
having the uniform brightness temperature and sharp outer edges
characteristic of subsonic expansion into the ambient intra-group
medium. The position angle of the much smaller ($l \sim 25$~kpc)
restarted central source is within $5^\circ$ of the main jets, ruling
out models that invoke jet re-orientation or two independent jets.
Compression and turbulence in the
backflows probably produce the irregular and { low polarization}
bright region behind the apex of each boomerang as well as several
{ features in the flow with bright heads and dark tails.}
\end{abstract}




\section{Introduction}

Most luminous and extended radio sources have a pair of collinear jets
thought to be aligned with the spin axis of the supermassive black
hole (SMBH) in the nucleus of the host galaxy \citep{bla77}. 
{ However,
$3$--$10$\% are X-shaped radio galaxies (XRGs) defined by having a
second set of jets or ``wings'' misaligned with the first
\citep{lea84, jos19}.}
The three main models for XRGs invoke (1) a sudden or
continuous reorientation of the nuclear SMBH spin axis
\citep[e.g.,][]{eke78, Kle95, den02}, (2) the superposition of two
independent linear jets produced by two SMBHs residing in the same
host galaxy \citep{lal05}, or (3) hydrodynamical backflows from the
over-pressured main jets deflected by the ellipsoidal hot interstellar
medium (ISM) of the host galaxy \citep[e.g.,][]{lea84, wor95, cap02,
  sar08}.

\begin{figure}
  \includegraphics[trim={1.4cm 0.5cm 1.6cm 0.5cm}, clip, 
    width=0.45\textwidth]{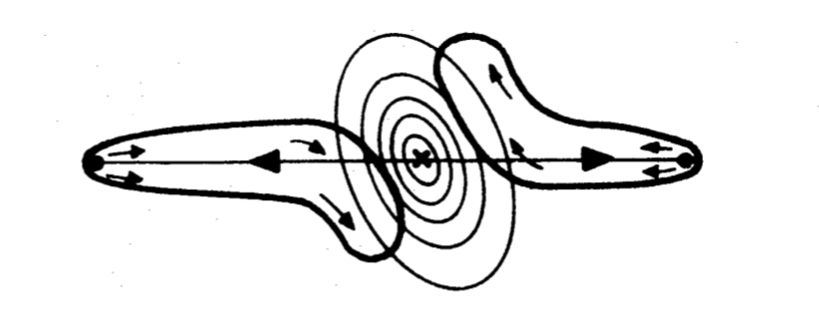}
  \caption{ This reproduction of fig.~6a in \citet{lea84} shows the
    main features of the hydrodynamical backflow model for XRGs.}
  \label{fig:leahy84}
\end{figure}

Fig.~\ref{fig:leahy84} is a reproduction of the prescient fig.~6a
in \citet{lea84} illustrating the hydrodynamical backflow model. The
nuclear SMBH ($\times$) emits two collinear radio jets (large
horizontal arrows) ending in hotspots (black dots). Hydrodynamical
backflows from the ends of the jets (small horizontal arrows)
initially preserve the axial symmetry of the jets. The axial symmetry
is broken by the oblique hot ISM of the host galaxy (thin elliptical
contours) deflecting the backflows in opposite directions (bent
arrows) and producing a radio source (heavy bent contours) with only
inversion symmetry about the nucleus. Although the radio source may
resemble a true XRG in low-resolution images, its actual shape is more
like a double boomerang.

In this paper we present and analyse new 1.28 GHz images of the giant
XRG PKS~2014$-$55 \citep{sar08,sar09} based on data from the recently
completed 64-element MeerKAT array of the South African Radio
Astronomy Observatory (SARAO) in the Northern Cape of South Africa.
{ We show that the morphology, spectrum and magnetic field
  structure of this source are very consistent with the hydrodynamical
  model and are inconsistent with other proposed models for X shaped
  sources.
}
{ The central component of PKS~2014$-$55 = PKS~2014$-$558 =
  PKS~J2018$-$556 \citep{wri90} is identified by position coincidence
  with the $m_v \approx 15.5$ Seyfert II elliptical galaxy PGC~064440
  \citep{pat89}. PGC~064440 has heliocentric redshift $z_\mathrm{h} =
  0.060629$ and velocity $v_\mathrm{h} = 18176 \pm 45
  \mathrm{~km~s}^{-1}$ \citep{jon09}; corrected to the cosmic
  microwave background (CMB) frame \citep{fix96} $z = 0.060252$ and $v
  = 18063 \pm 46 \mathrm{~km~s}^{-1}$. All absolute quantities in
  this paper were calculated for a $\Lambda$CDM universe with $H_0 =
  70 \mathrm{~km~s}^{-1} \mathrm{~Mpc}^{-1}$ and $\Omega_\mathrm{m} =
  0.3$ using equations in \citet{con18}. Thus PGC~064440 is at
  comoving distance $D_\mathrm{C} \approx 254$~Mpc,
(bolometric) luminosity distance $D_\mathrm{L} \approx 270$~Mpc,
and angular diameter distance $D_\mathrm{A} \approx 240$~Mpc
so $1\arcmin \approx 70\,\mathrm{kpc}$.}

The radio observations and data reduction are described in
Section~\ref{sec:obs}, and the resulting images are presented in
Section~\ref{sec:images}. Section~\ref{sec:galaxy} discusses the
morphology of the radio source and its host galaxy.
The data are interpreted with the asymmetric hydrodynamical model in
Section~\ref{sec:model}. Our results are summarized in
Section~\ref{sec:summary}.

\section{Observations and Data Reduction}\label{sec:obs}
  
The XRG PKS~2014$-$55 and its unpolarized gain/\allowbreak
bandpass/flux-density calibrator PKS~B1934$-$638 were observed by
MeerKAT for 9.7~h on 2019 October 11 using 55 of the 64 13.5\,m
diameter antennas in the array. One 10~min scan on the polarization
calibrator 3C~138 was also included. Additional information about
MeerKAT and its specifications can be found in \citet{jon16},
\citet{cam18}, and \citet{mau20}. The maximum baseline length used was
nearly 8~km, giving $\theta \approx 7\,\farcs4$ FWHM resolution at
$\nu = 1.28$~GHz. All four correlations XX, YY, XY, and YX of the
orthogonal linearly polarized feeds were divided into 4096 spectral
channels of width 0.208984~MHz. The 856~MHz total bandpass centred on
1.284~GHz includes the 1.420\,GHz \ion{H}{i} line near the $z = 0.06$
host galaxy PGC 064440 redshifted to 1.34\,GHz, where the channel
width is $\Delta v \approx 47 \mathrm{\,km\,s}^{-1}$. The averaging
time was 8~s.

\subsection{Continuum Flagging and Calibration\label{ContCal}}

The $(u,v)$ data were converted from the archive format to AIPS format
using MeerKAT's {\small KATDAL}
package\footnote{\url{https://libraries.io/github/ska-sa/katdal}}.
The initial radio-frequency interference (RFI) flagging followed that
by \citet{mau20}. We trimmed 144 channels from each end of the
bandpass and merged the 3808 remaining spectral channels into eight
spectral windows. Subsequent editing and calibration used the {\small
  OBIT} package
\citep{cotton08}\footnote{\url{http://www.cv.nrao.edu/~bcotton/Obit.html}}.
In each calibration step, deviant solutions were detected and flagged
along with the corresponding $(u,v)$ data. The
gain/bandpass/flux-density calibrator is essentially unpolarized, so
the approximation XX=YY is valid. Standard structural and spectral
models for PKS~B1934$-$638 and 3C~138 were used as appropriate. Our
flux-density scale is based on the \citet{rey94} polynomial fit for
the spectrum of PKS~B1934$-$638:
\begin{eqnarray}
  \log(S) = -30.7667 + 26.4908 \log\biggl(\frac{\nu}{\mathrm{MHz}}\biggr) ~~~ \nonumber \\
  - 7.0977 \log\biggl(\frac{\nu}{\mathrm{MHz}}\biggr)^2
  +0.605334 \log\biggl(\frac{\nu}{\mathrm{MHz}}\biggr)^3~,
\end{eqnarray}
where $S$ is the flux density in Jy and $\nu$ is the frequency.

The main flagging and calibration steps were:
\begin{enumerate}
\item{\bf Fixed flagging:} Frequency ranges known to contain strong,
  persistent RFI were flagged. Projected baselines shorter than the
  13.5\,m dish diameter were flagged to eliminate shadowing.
\item{\bf Hanning smoothing:} The spectral channels were Hanning smoothed
  (channel weights $1/4,\, 1/2,\,\allowbreak 1/4$) to reduce possible
  Gibbs ringing from strong narrowband RFI. This doubled the
  effective channel bandwidth.
\item{\bf Initial flagging:} Running medians in time and frequency
  were compared with the data to reveal variable and/or narrowband RFI
  for flagging.
\item{\bf Initial X-Y phase calibration:} Cross-hand phase corrections
  were determined from the noise diode calibration signals injected
  into each data stream at the beginning of the observing session.
\item{\bf Delay calibration:} Residual group delays were determined
  from PKS~B1934$-$638.
\item{\bf Bandpass calibration:} Amplitude and phase correction
  spectra were determined from  PKS~B1934$-$638.
\item{\bf Amplitude and phase calibration:} Complex gain solutions for
  PKS~B1934$-$638 were determined and applied to the target field. 
\item{\bf Flagging of calibrated data:} Flagging operations for which
  calibrated data are needed were done.
\item{\bf Repeat:}  Flags from the steps i--viii were kept and
  the calibration steps iv--viii were repeated.
\item{\bf Polarization calibration:} After a further averaging of two
  spectral channels, instrumental polarization was determined from the
  unpolarized calibrator PKS~B1934$-$638; solutions were obtained in
  14.2\,MHz blocks. The cross-hand delay and phase were determined
  from the polarized calibrator 3C~138.
{ All polarization calibration parameters, including feed
  orientation, were determined jointly in a nonlinear least-squares
  solution using all calibrators.} 
\end{enumerate}
Finally, the calibrated $(u,v)$ data were averaged in time to reduce
their volume. The averaging times were subject to the
baseline-dependent constraint that averaging reduce the amplitudes by
$\leq 1$\% inside the circle of radius $\rho = 1\fdg2$ centred on the
target, and they never exceeded 30\,s.

\subsection{Continuum Imaging}\label{ContImage}

The continuum imaging used the Obit task MFImage, which was described
in more detail by \citet{cot18}. The non-coplanar array in our
extended synthesis was corrected by covering the imaged sky with small
tangential facets out to radius $\rho= 1\fdg2$ and placing additional
facets on outlying sources stronger than $S \approx 6 \mathrm{\,mJy}$
from the 843~MHz SUMSS catalog \citep{mau03}. Variations of sky
brightness and antenna gain across our wide bandpass were accommodated
by dividing the observed spectrum into 34 frequency bins having
$\sim2$\% fractional bandwidths. The frequency bins were imaged
independently and CLEANed jointly. Three iterations of phase
self-calibration were applied. The CLEAN window was mostly generated
automatically, but with some manual assistance to cover all of the
extended emission. The multi-resolution Stokes I CLEAN used 1,032,004
components, a 3\% loop gain, and included 3.394 Jy of total flux
density after CLEANing to a depth of $15\,\mu\mathrm{Jy\,beam}^{-1}$.
Spectra were least-squares fitted in each pixel using frequency bin
weights inversely proportional to the square of the image-plane rms
noise.

Stokes Q and U were imaged out to a radius $\rho = 1\fdg0$ and CLEANed
for 50,000 components to a depth of $28\,\mu\mathrm{Jy\,beam}^{-1}$ in
Stokes Q and $15\,\mu\mathrm{Jy\,beam}^{-1}$ in Stokes U. Rotation
measures (RMs) and electric-vector polarization angles (EVPAs)
corrected to zero wavelength for each pixel were derived by a search
in RM space, { essentially taking the peak of the rotation measure
  synthesis function \cite{Bre05}.} The RM that gave the highest polarized
intensity in the average RM-corrected value of the $P =\sqrt{Q^2 +
  U^2}$ was taken as the RM, the peak average $P$ as the polarized
intensity, and the polarization angle of the RM-corrected average
$Q\ +\ iU$ as the EVPA at zero wavelength. The EVPA at zero wavelength
is orthogonal to the source magnetic field vector \emph{$B$}
{ integrated through the source.}
{ There is little evidence for effects beyond a simple
  well-resolved, external Faraday screen. The fractional polarization
  image was derived by first correcting the frequency averaged
  polarized intensity for the Ricean bias and divided by the frequency
  averaged total intensity.}

\subsection{Continuum Spectral Index\label{SpectralIndex}}
{ The imaging described in Section~\ref{ContImage} used tapering in the
  outer portion of the $(u,v)$ plane to minimize the variation of
  resolution with frequency but did not address a similar problem in
  the inner portion of the $(u,v)$ plane that distorts the
  spectral-index image. Further imaging similar to that described in
  Section~\ref{ContImage} was done using a Gaussian taper in the inner
  $(u,v)$ plane with rms length $\sigma = 500\lambda$ to reduce the
  frequency dependence of the short-baseline coverage.
  In order to better explore the faintest emission,
  we convolved the images of the individual frequency planes to $\theta_{1/2}
  = 15\arcsec$ before making frequency-dependent primary beam
  corrections and fitting for the spectral index. }

\subsection{Spectral-line Calibration and Imaging}

The 1420\,MHz \ion{H}{i} line frequency at the redshift $z = 0.06$ of
PGC~064440 is 1340 MHz. We reduced a $50\,\mathrm{MHz}$ subset of the
visibility data including this frequency using the IDIA
Pipeline\footnote{https://www.idia-pipelines.github.io} to produce a
spectroscopic data cube containing the \ion{H}{i} line around
PGC~064440. The IDIA pipeline is based entirely on {\small CASA}
\citep{mcm07} tasks. The basic workflow follows the calibration
process described above. After the continuum subtraction using {\small
  CASA}'s  UVCONTSUB task to fit a polynomial to the
visibility spectra, we used {\small CASA}'s TCLEAN {\bf task}
with the widefield gridder performing 1000 CLEAN iterations per
channel. A Briggs weighting with $\mathrm{robust} = 0.5$ was used to
optimise the shape of the dirty beam while minimizing sensitivity
loss. 
We  measured the image noise to be $\sigma\sim125\,\mathrm{\mu
  Jy\,beam^{-1}}$.
The FWHM resolution in the \ion{H}{i} cube is
$\theta=\,19\,\farcs2 \times 17\,\farcs6$.

\section{The Radio Images}\label{sec:images}

\subsection{Total Intensity Continuum}

\begin{figure*}
  \includegraphics[trim={1.5cm 3.cm 1.5cm 3cm}, clip, 
    width=\textwidth]{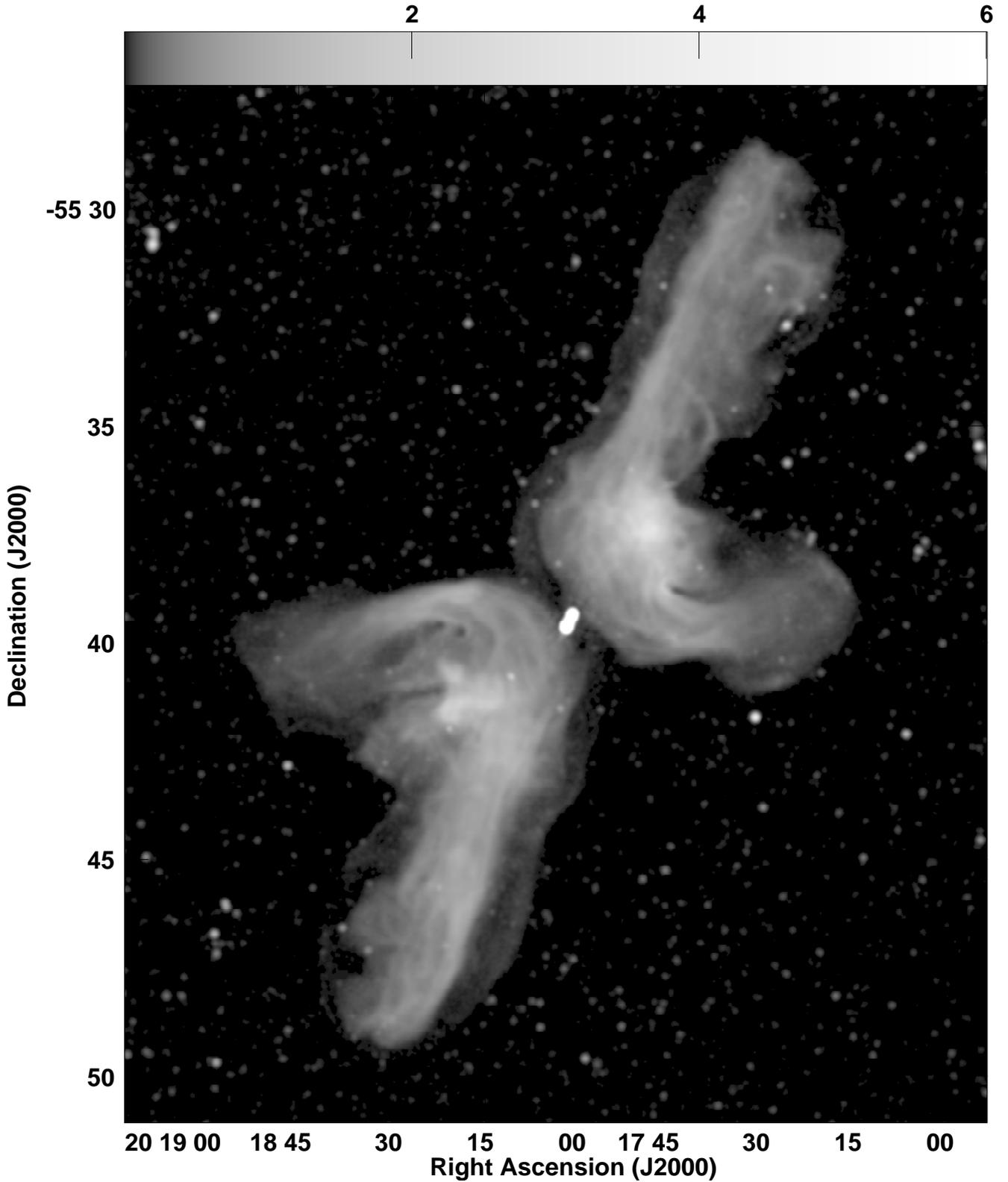}
  \caption{ This 1.28 GHz MeerKAT ``sky'' image of PKS~2014$-$55 was
    restored with a $\theta_{1/2} \approx 7\,\farcs4 \approx 8.6
    \mathrm{~kpc}$ FWHM Gaussian beam, corrected for the $-4.6 \pm 1.0
    \,\mu\mathrm{Jy\,beam}^{-1}$ bowl, and divided by the primary
    attenuation pattern. The rms noise plus confusion is $\sigma = 3.5
    \pm 0.2 \,\mu\mathrm{Jy\,beam}^{-1} = 48 \pm 3 \mathrm{\,mK}$.
    The logarithmic grey scale spans
    $0.006$--$6\,\mathrm{mJy\,beam}^{-1}$ { (grey-scale bar at
      top)} to reveal the extremely faint ($T_\mathrm{b} \sim 0.5
    \mathrm{\,K}$) radio cocoons
    { most conspicuous on the east side of the northern arm and on
      the west side of the southern arm.}  The bright central region
    is completely saturated, but it is accurately represented by the
    contour map in Fig.~\ref{fig:xcore}.}
  \label{fig:ipol}
\end{figure*}

UnCLEANed flux from the very extended radio source PKS~2014$-$55 and
background sources in the primary beam, combined with the lack of
projected baselines shorter than the 13.5\,m antenna diameter, left a
wide but
{ very} shallow negative ``bowl'' in the total-intensity image. We
used the {\small AIPS} task {\small IMEAN} to measure the mode of the
intensity { distribution} of pixels in source-free areas near
PKS~2014$-$55; it is $-4.6 \pm 1.0 \,\mu\mathrm{Jy\,beam}^{-1}$. To
fill in the bowl, we added $4.6 \,\mu\mathrm{Jy\,beam}^{-1}$ to the
image zero level.
{ We divided this image by the circularised $67\arcmin$ FWHM
  primary beam attenuation pattern specified by \citet{mau20}
  equations 3 and 4 to yield the final ``sky'' image shown in
  Fig.~\ref{fig:ipol}. The actual primary beam is slightly elliptical
  with axial ratio $a/b \approx 1.04$ and rotates with parallactic
  angle on the sky. However, the maximum attenuation error introduced
  by our circular approximation is a negligible $<0.3$\% even at the
  ends of PKS 2014$-$55, $\rho = 12\arcmin$ from the pointing centre.
}

{ The image} was restored with a $7\,\farcs43 \times 7\,\farcs35$
Gaussian beam, so
the brightness temperature corresponding to a spectral intensity
$S_\mathrm{p} = 1 \,\mu\mathrm{Jy\,beam}^{-1}$ is $T_\mathrm{b}
\approx 13.7 \mathrm{~mK}$ at $\nu = 1.28 \mathrm{~GHz}$. The rms
noise plus confusion is $\sigma = 3.5 \pm 0.2
\,\mu\mathrm{Jy\,beam}^{-1}$.

We used the {\small AIPS} task {\small TVSTAT} to measure the total
1.28~GHz flux density in the $\Omega = 1.02 \times 10^{-5}
\mathrm{\,sr}$ region bounded by the faint ``cocoon'' that makes
Fig.~\ref{fig:ipol} resemble the photograph of a jellyfish; it is $S =
2.36 \pm 0.08 \mathrm{~Jy}$, with the quoted error being dominated by
an estimated 3\% uncertainty in the absolute flux-density
scale. Inside the region 
subtended by
PKS~2014$-$55, the $4.6 \times 10^{-6} \mathrm{\,Jy\,beam}^{-1}$ bowl
correction contributed only $S \approx 0.03$~Jy
to the total flux density.
{ The smooth bowl correction is only $\sim12$\% of the
  average cocoon brightness, so our detection of the sharp-edge
  cocoons is quite reliable.}
{ Most published flux densities \citep[e.g.][]{wri90,hin14} at
  nearby frequencies appear to be about 25\% lower than ours, perhaps
  because they didn't capture the full extent of this very large
  source.}  The Murchison Widefield Array (MWA) flux densities
\citep{hin14} $S = 15.2 \pm 0.8$, $13.3 \pm 0.7$, $11.3 \pm 0.6$, and
$8.8 \pm 0.5$~Jy at 120, 149, 180, and 226\,MHz, respectively, are
consistent with PKS~2014$-$55 having an overall spectral index $\alpha
\equiv + d \ln S / d \ln \nu$ in the range $-1.0 < \alpha < -0.8$.

\begin{figure}
  \includegraphics[trim={1.7cm 5.cm 1.5cm 5cm}, clip, 
    width=1.\columnwidth]{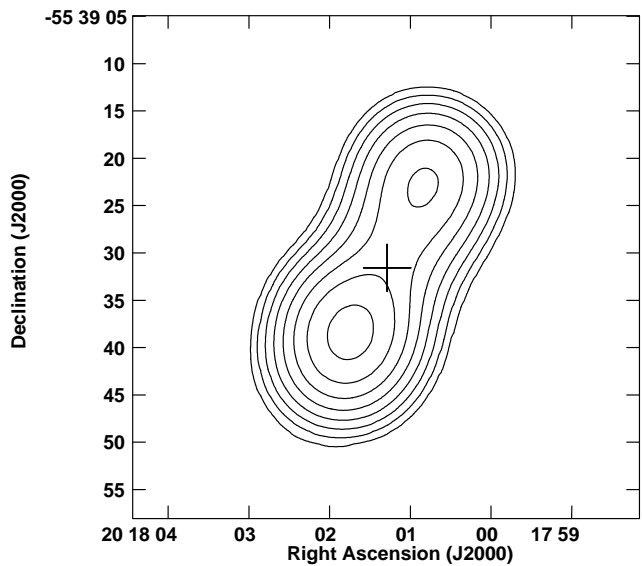}
  \caption{ The bright central region of PKS~2014$-$55 at 1.28~GHz is
    represented by contours at $1\mathrm{~mJy\,beam}^{-1} \times 2^0,
    \,2^1,\, 2^2, \dots, 2^7$.
       { The cross marks the accurate position
    J2000 $\alpha = 20^\mathrm{h}~18^\mathrm{m}~ 01\fs286$, $\delta =
    -55^\circ~39\arcmin~31\farcs59$ \citep{skr06} of the near-infrared
    source 2MASX J201801.27$-$553931.5 identified with PGC~064440.}}
  \label{fig:xcore}
\end{figure}

The central region in our 1.28~GHz total-intensity image of PKS~2014$-$55
(Fig.~\ref{fig:xcore}) is well fit by the sum of three Gaussian
components: two with completely free size and position parameters plus
a (forced) point source with free position and flux density
representing a possible radio core between them. The results are
listed in Table~\ref{tab:gfits}. Although it is not clearly visible
in Fig.~\ref{fig:xcore}, the $32 \pm 6$~mJy core component is
required for the best fit and its fitted position is $<1\arcsec$ from
the nucleus of the host galaxy PGC~064440. The line connecting the
two outer components has position angle $PA = 154^\circ$ measured from
north to east.

\begin{table}
\centering
\caption{The positions, 1.28~GHz flux densities, deconvolved major and
  minor diameters between half-maximum points, and major-axis position
  angles of Gaussian fits to the central region of PKS~2014$-$55.}
  \label{tab:gfits}
  \begin{tabular}{c c c c c}
    \hline
    \multicolumn{2}{c}{J2000} & $S$~ & $\phi_\mathrm{M} \times \phi_\mathrm{m}$ &
    $PA$ \\
    $\alpha$ & $\delta$ & (mJy) & ($\arcsec \times \arcsec$)  &
    \llap{(}deg\rlap{)} \\
    \hline
    $20^\mathrm{h}\,18^\mathrm{m}\,00\,\fs84$ &
    $-55^\circ~39\arcmin~22\,\farcs9$ &
    88 & $5.2 \times 1.4$ & 159 \\
    $20^\mathrm{h}\,18^\mathrm{m}\,01\,\fs2\hphantom{3}$ &
    $-55^\circ~39\arcmin~31\arcsec\hphantom{.1}$ & 
    32\rlap{$\pm 6$} & $0 \times 0$ & \dots \\
    $20^\mathrm{h}\,18^\mathrm{m}\,01\,\fs74$ &
    $-55^\circ~39\arcmin~38\,\farcs5$ &
    \llap{2}11 & $5.4 \times 1.1$ & 154 \\
    \hline
  \end{tabular}
\end{table}

\citet{sar08} imaged the central region of PKS~2014$-$55 with
sub-arcsec resolution and found five nearly collinear radio
components: a central core, an inner double source, and an outer
double source. The position angles of their inner and outer doubles
are $+150^\circ$ and $+156^\circ$, respectively. Thus each long,
narrow Gaussian component in Table~\ref{tab:gfits} is a blend of
two relatively compact sources.
The \citet{sar08} radio core has a fairly flat ($\alpha > -0.5$)
spectrum above $\nu = 1.28 \mathrm{\,GHz}$, so it is probably
synchrotron self-absorbed and completely unresolved.

\subsection{Polarization}

\begin{figure*}
 \includegraphics[trim={1.5cm 3.0cm 1.5cm 3.0cm}, clip, 
    width=\textwidth]{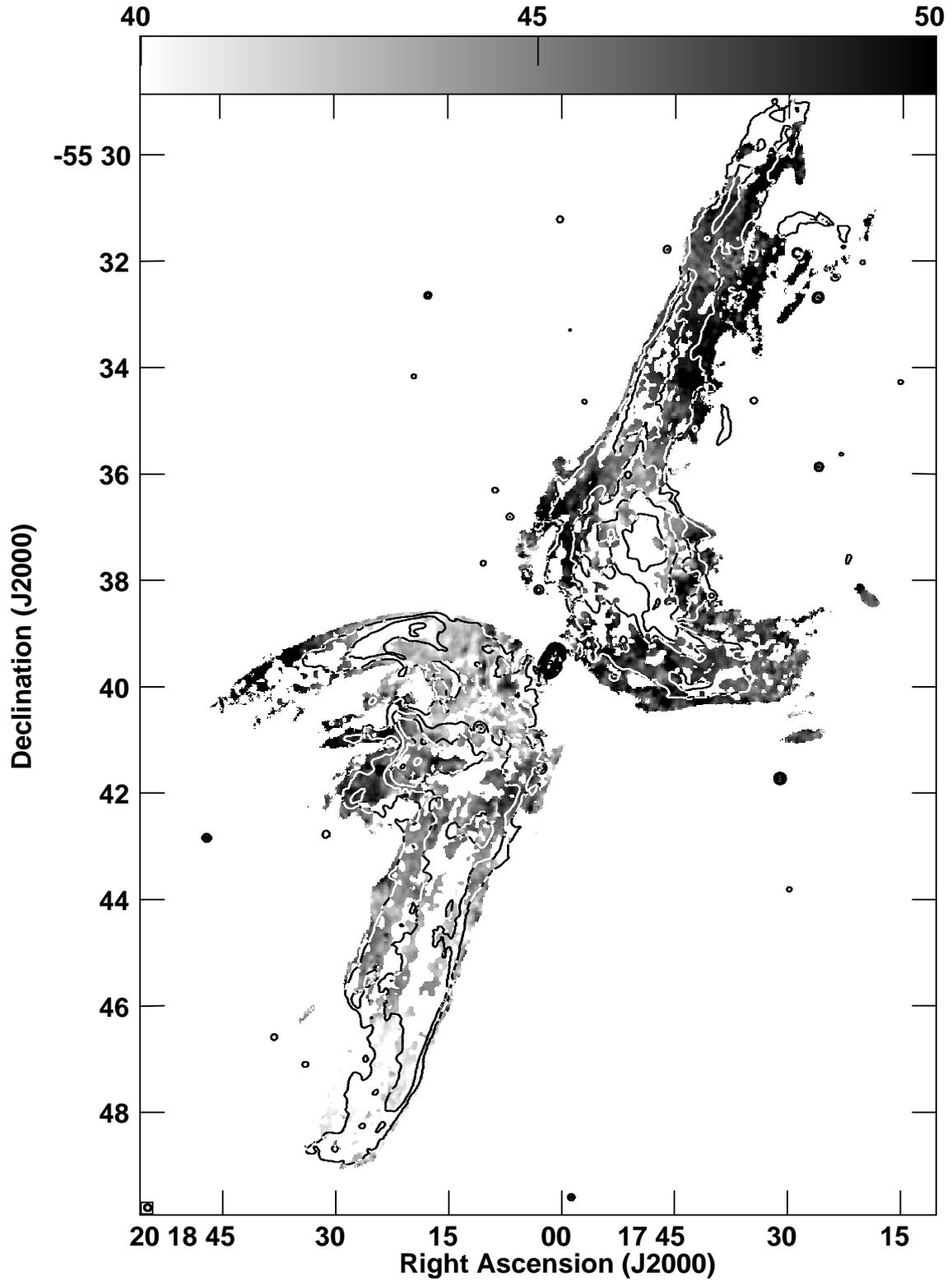}
 \caption{This image shows the rotation measure RM on a
   grey scale. The scale bar at the top has units
   $\mathrm{rad~m}^{-2}$. Total-intensity contours are plotted at
   brightnesses $100\,\mu\mathrm{Jy~beam}^{-1} \times2^0,~2^1,~2^2,
   \dots$. The $\theta_{1/2} \approx 7\,\farcs4$ restoring beam half-power
  circle is shown in the small box in the lower left corner.
}
\label{RMfig}
\end{figure*}

The rotation measure in front of PKS~2014$-$55
{ is shown in Figure \ref{RMfig}. It}
is $\mathrm{RM} \sim 40
\mathrm{\,rad\,m}^{-2}$ and varies by only a few rad m$^{-2}$ across
the source, so it may originate primarily in our Galaxy.

\begin{figure*}
  \centering
  \includegraphics[trim={0.9cm 2.5cm 1.5cm 4.3cm}, clip, 
   width=1.0\textwidth]
 {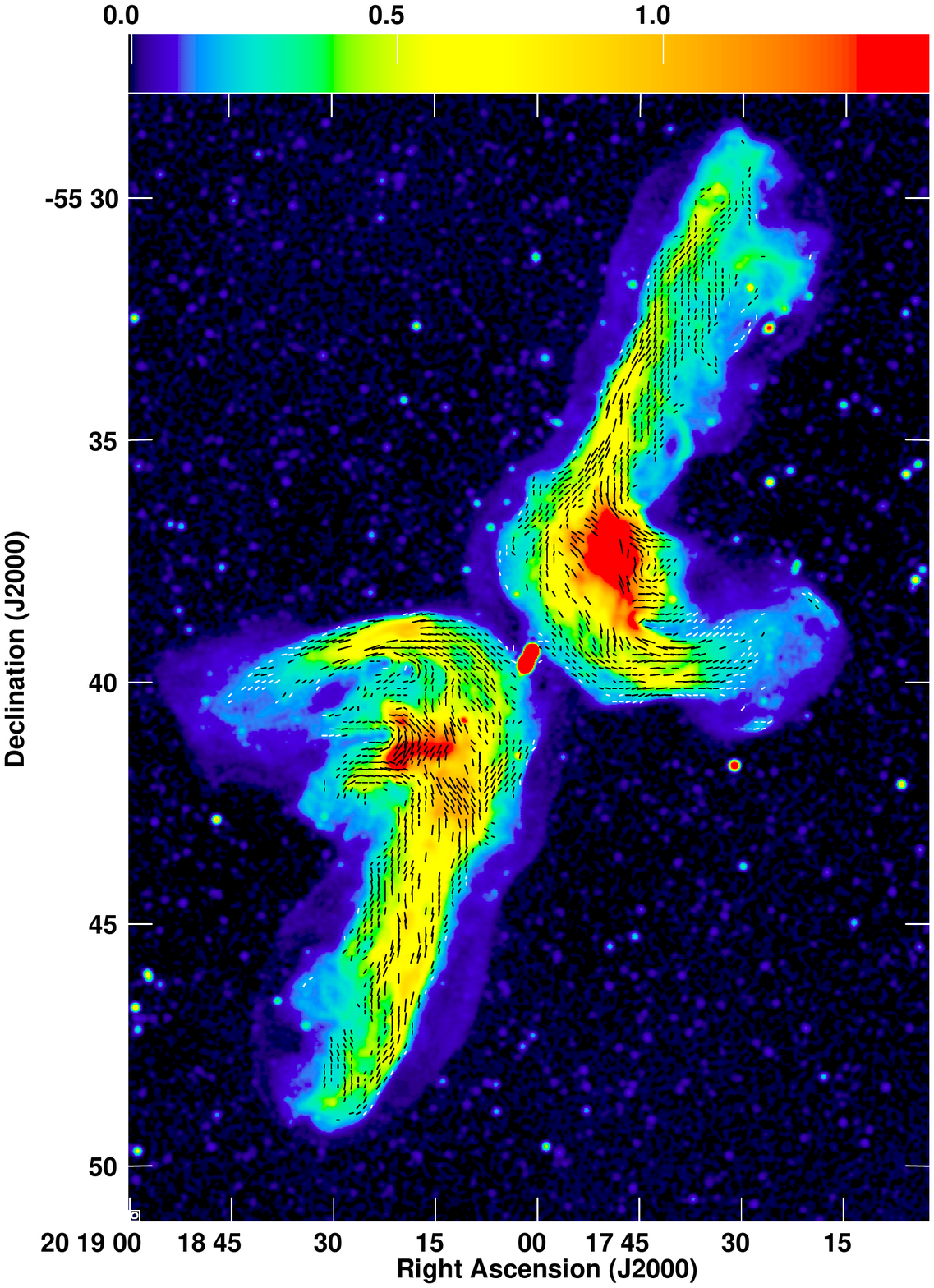}
 \caption{The black and white lines are parallel to
   \emph{$B$} and have length $1\arcsec$ per $P = ${ 4.2\%
   fractional polarization}. The colour bar gives
   the linear total-intensity scale between $-0.005$ and
   $1.5\,\mathrm{mJy\,beam}^{-1}$. The $\theta_{1/2} \approx 7\,\farcs
   4$ restoring beam half-power circle is shown in the small box in
   the lower left corner.}
\label{fig:poln}
\end{figure*}

Polarization ``B'' vectors at zero wavelength parallel to the magnetic
field orientation in the emitting region are plotted as short line
segments in Fig.~\ref{fig:poln}.
{ The lengths of the vectors are proportional to the fractional
polarization.}
The magnetic vectors are nearly
parallel to the main jets, as in most FR\,II \citep{fan74} sources
\citep{bri94}. In contrast, the magnetic vectors are usually
perpendicular to the jets in FR\,I sources \citep{bri84}.
Fig.~\ref{fig:poln} shows that the magnetic field \emph{$B$} closely
follows the apparent flow around the bends in PKS~2014$-$55.

{ The fractional polarization is high (30--50\%) over most of the
  main lobes indicating very organized magnetic fields.
In the secondary lobes, the fractional polarization approaches 80\%
indicating very little scatter in the magnetic field orientations.
On the other hand, the brighter Stokes I regions at the apexes of the
``boomerangs'' typically have around 15\% fractional polarization
indicating a more tangled magnetic field structure.
}

{
\subsection{Continuum Spectral Index}
The spectral-index image is given in Figure~\ref{fig:SI}. 
The spectrum in
the lobes is steep everywhere, flattening somewhat near the bright
regions inside the apexes of the ``boomerangs'' and becoming very
steep in the cocoons and near the ends of the secondary lobes. We
hesitate to provide quantitative estimates of the spectral-index
errors because they are largely systematic, caused by limited sampling
near the centre of the $(u,v)$ plane.
Further analysis of the spectral index is being developed for a
subsequent paper.
\begin{figure*}
  \centering
  \includegraphics[trim={1.9cm 3.0cm 0.5cm 4.3cm}, clip, 
   width=1.0\textwidth]
 {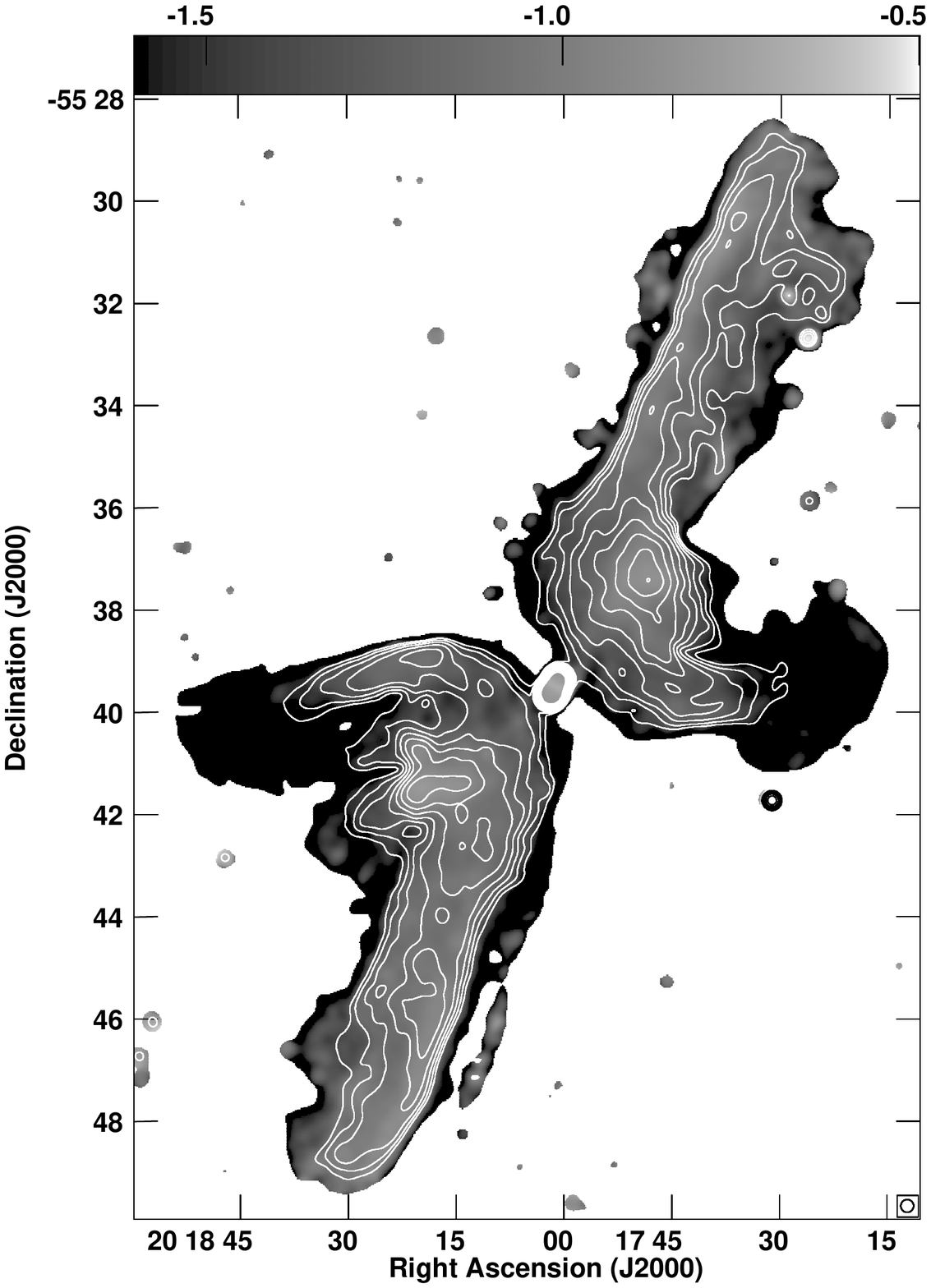}
 \caption{The grey scale image shows the spectral index $\alpha$ 
   with values given  by the scale bar at the top. The white
   contours mark total intensities
   $0.5\,\mathrm{mJy\,beam}^{-1} \times 2^0,\,2^{1/2},\,2^1$\dots
   The $15\arcsec$ half-power circle is drawn inside the small box
   at the lower right.}
\label{fig:SI}
\end{figure*}
}

\section{The Radio Source PKS~2014$-$55 and its Host Galaxy PGC~064440}
\label{sec:galaxy}

\subsection{The Radio Source PKS~2014$-$55}\label{sec:radiosource}

The $S = 2.36\, \mathrm{Jy}$ radio source with spectral index
$\alpha \approx -0.8$ has spectral luminosity $L_\nu \approx 2.0
\times 10^{25} \,\mathrm{W~Hz}^{-1}$
at $\nu = 1.28$~GHz in the source frame. Such luminous sources
\citep{led96} usually have FR\,II morphologies \citep{fan74}
characterised by narrow jets leading to hotspots in edge-brightened
lobes. PKS~2014$-$55 does not (Fig.~\ref{fig:ipol}); its long,
filamentary, and diffuse main ``jets'' are only decaying relics of
activity that ceased millions of years ago.
{ \citet{sar09} found several examples of XRGs lacking hotspots at
  the ends of their relic jets and noted that the current lack of
  hotspots cannot be used to rule out backflows from earlier jets that
  have since decayed.}

The current spectral brightnesses of the relic jets are only
$T_\mathrm{b} \sim 14\,\mathrm{K}$, so their
minimum-energy magnetic field strengths $B_\mathrm{min}$
{ \citep[e.g.,][]{pac70, wor06} are low. For electrons emitting at
  their critical frequencies from $\nu_1 = 10^7\,\mathrm{Hz}$
  to $\nu_2 = 10^{10}\,\mathrm{Hz}$, proton/electron energy ratio $\kappa$,
  and source line-of-sight depth $d$,
\begin{equation}\label{eqn:bmintunits}
  \biggl( \frac{B_\mathrm{min}}{\mu\mathrm{G}} \biggr) \approx 0.57
  \Biggl[\biggl( \frac{T_\mathrm{b}}{\mathrm{K}} \biggr)
       \biggl( \frac{\nu(1+z)}{\mathrm{GHz}} \biggr)^{2-\alpha}
      \biggl( \frac{\mathrm{kpc}}{d} \biggr)
    \Biggr]^{2/7} (1+\kappa)^{2/7}
\end{equation}
in convenient astronomical units.  } Even for a line-of-sight depth $d
= 100 \mathrm{\,kpc}$ and a high proton/electron energy ratio $\kappa
= 2000$, the magnetic field strength is only $B_\mathrm{min} \sim 3
\,\mu\mathrm{G}$ and the corresponding synchrotron lifetime $\tau \sim
c_{12}B^{-3/2} \sim 1.3 \times 10^{16} \mathrm{\,s} \sim 4 \times 10^8
\mathrm{\,yr}$ \citep{pac70} is very long. The energy density of the
CMB at $z = 0.06$ is the same as that of a $B \sim 3.7\,\mu\mathrm{G}$
magnetic field, so inverse--Compton (IC) scattering off the CMB
reduces the radiative lifetimes in the relic jets to $\tau \sim 2
\times 10^8 \mathrm{\,yr}$.
{ The spectral steepening at the ends of the wings visible in
  Figure~\ref{fig:SI} indicate ages $\tau\gtrsim 10^8$\,yr. This
  result is typical of giant radio galaxies \citep{ish99}.}  The
minimum relativistic pressure
\begin{equation}\label{eqn:pmin}
 P_\mathrm{min} = \frac{13 B_\mathrm{min}^2}{72 \pi}
\end{equation}
in most of
PKS~2014$-$55  is $P_\mathrm{min} \sim
10^{-14} (1 + \kappa)^{4/7} \mathrm{\,dyne\,cm}^{-2}$.

At $D_\mathrm{A} \approx 240$~Mpc, $1\arcsec \approx 1.164
\mathrm{~kpc}$ and the largest angular extent $\phi \approx 22\farcm5$
of PKS~2014$-$55 implies a projected overall length $l \approx
1.57$~Mpc. This is more than twice the traditional minimum size
defining a giant radio source, $l \approx 1$~Mpc for $H_0 = 50
\mathrm{~km~s}^{-1} \mathrm{~Mpc}^{-1}$ \citep{wil74}, or $l \approx
0.7$~Mpc for $H_0 = 70 \mathrm{~km~s}^{-1} \mathrm{~Mpc}^{-1}$. Even
the backflow wings of PKS~2014$-$55 easily satisfy this criterion:
their total projected extent is $\phi \approx 14\farcm0$ or $l \approx
0.98$~Mpc. The two { long arms} have nearly equal projected lengths
($11\farcm7$ and $10\farcm8$ for the NW and SE { arms},
respectively) and flux densities (1.033 and 1.025~Jy), so they show no
evidence for relativistic time delays or flux boosting.

PKS~2014$-$55 extends far beyond the virial radius of its host galaxy
and directly probes the ambient intergalactic medium (IGM). PGC
064440 is not in a cluster environment rich enough to have a significant
intracluster medium. \citet{mal15} made a spectroscopic study of the
Mpc-scale environments for a sample of low-redshift giant radio
galaxies including PKS~2014$-$55. The number density of galaxies more
luminous than $-19.49$~mag in a cylinder of 1~Mpc radius and 24~Mpc
length along the line-of-sight centred on PKS~2014$-$55 is only $n
\sim 0.066 \mathrm{~Mpc}^{-3}$, a typical density in galaxy groups
and poor clusters, but a factor of 10 lower than in galaxy clusters.

\begin{figure*}
  \centering
  \includegraphics[trim={0.5cm 0.5cm 0.5cm 1cm}, clip, 
    width = 0.49\textwidth]{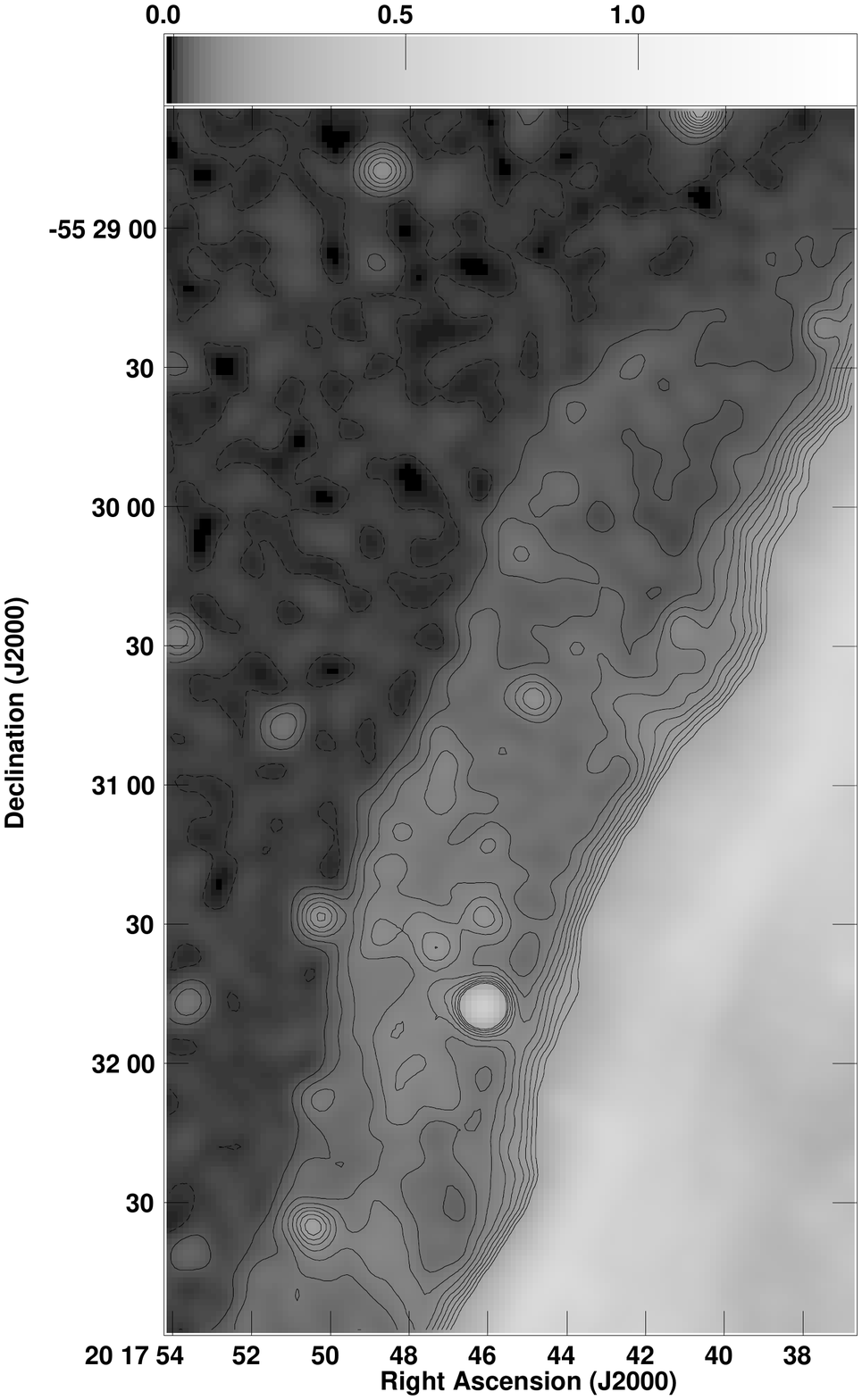}
  \includegraphics[trim={0.5cm 0.5cm 0.5cm 1cm}, clip,
    width= 0.49\textwidth]{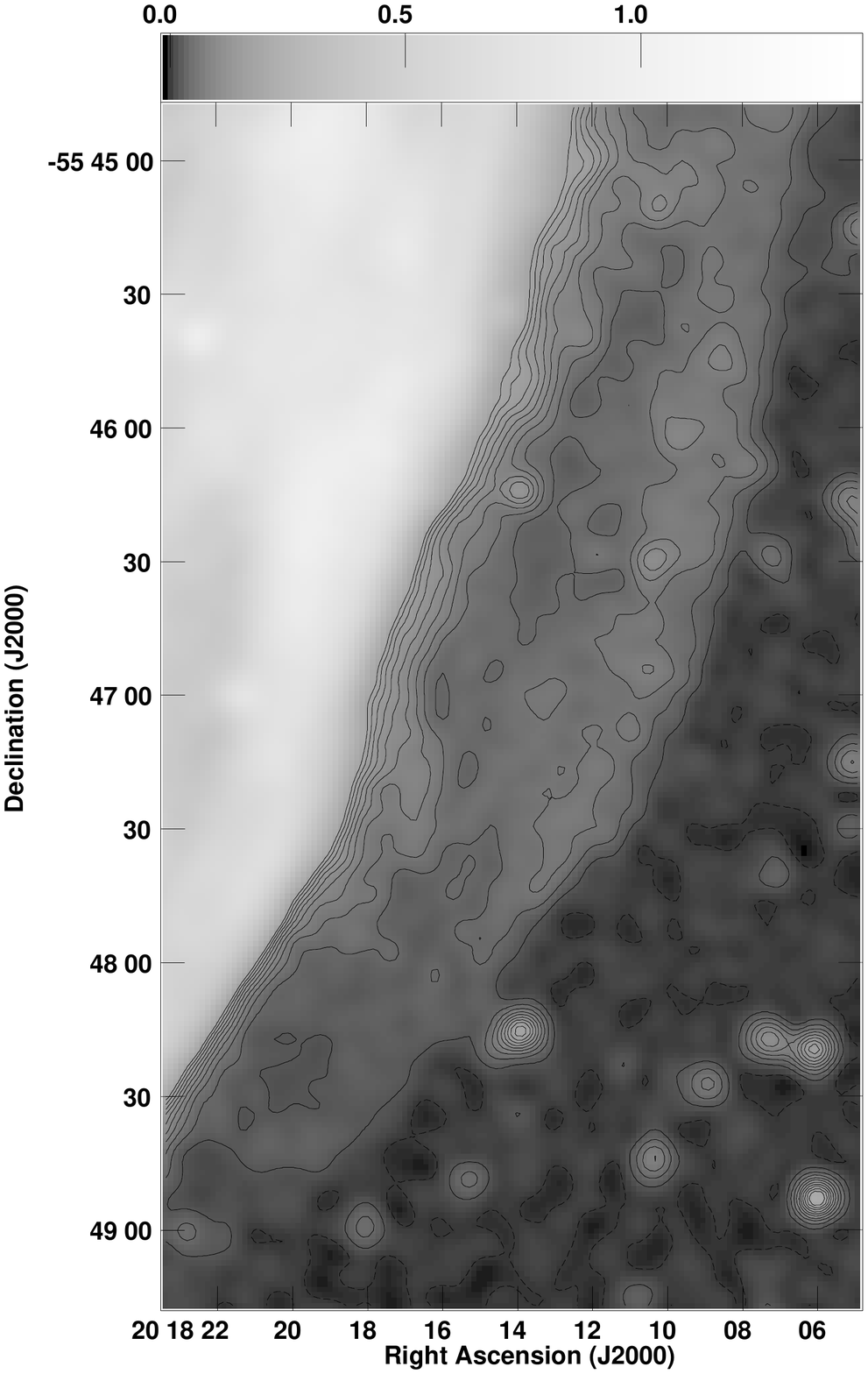}
  \caption{Portions of the cocoons surrounding the northern  (left
    panel) and the southern { relic} jets (right panel) are shown with contours
    at $T_\mathrm{b} = 0.25\,\mathrm{K} \times 1, \, 2,\, 3,\, \dots,
    10$ to emphasize their sharp inner and outer boundaries. The
    average brightness temperature in the cocoons is $\langle
    T_\mathrm{b}\rangle \sim 0.5$~K. The logarithmic grey scale spans
    $-0.12 \mathrm{\,K} = -0.008 \mathrm{\,mJy\,beam}^{-1}$ (black) to
    $+20 \mathrm{\,K} = +1.46 \mathrm{\,mJy\,beam}^{-1}$ (white)
    { as indicated by the scale bar at the top.}}
\label{fig:cocoon}
\end{figure*}

{ The faint radio cocoons in Fig.~\ref{fig:cocoon} are defined by
  their} fairly constant brightness temperatures $T_\mathrm{b} \sim
0.5$~K between sharp inner and outer boundaries.  {
  Figure~\ref{fig:SI} shows that they have the steep spectra produced
  by radiative losses, so, like the relic jets, they too may be relics
  of even earlier activity.}  
Inserting $T_\mathrm{b} = 0.5$~K, $\nu =
1.28$~GHz, and line-of-sight depth $d = 100$~kpc into
Equation~\ref{eqn:bmintunits} yields $B_\mathrm{min} \approx 0.15
(1 + \kappa)^{2/7} \,\mu\mathrm{G}$
in the cocoon of PKS~2014$-$55. The corresponding magnetic energy
density is $U_B \sim 1.0 \times 10^{-15} (1+\kappa)^{4/7} \,
\mathrm{erg~cm}^{-3}$
and the minimum relativistic pressure in the cocoons is
$P_\mathrm{min} \sim 1.3 \times 10^{-15} (1+\kappa)^{4/7}
\,\mathrm{dyne~cm}^{-2}$.
These low-pressure cocoons are exceptionally sensitive barometers for
measuring the pressure of the intra-group medium (IGrM)  or the IGM
\citep{mal13}.

If the pressure $P_\mathrm{e}$ in the external medium is less than the
cocoon pressure $P_\mathrm{c}$, the cocoon should expand laterally with
speed $v_\perp$ at which the ram pressure balances the cocoon
overpressure:
\begin{equation}\label{eqn:ram}
  v_\perp = \biggl( \frac {P_\mathrm{c} - P_\mathrm{e}}
  {\rho_\mathrm{e}} \biggr)^{1/2}~,
\end{equation}
where $\rho_\mathrm{e}$ is the external mass density. The IGM may
contain half of all baryons and $\Omega_\mathrm{b} \approx 0.046$, so
a lower limit to the external mass density at redshift $z = 0.06$ is
the mean baryonic IGM density
\begin{equation}\label{eqn:whimrho}
  \rho_\mathrm{e} \ga \biggl( \frac{\Omega_\mathrm{b}}{2} \biggr)
  \frac {3H_0^2 (1+z)^3}{ 8 \pi G}
  \approx 2.5 \times 10^{-31} \mathrm{~g~cm}^{-3}~.
\end{equation}
For a primordial abundance of fully ionized H and He, the mean mass
per particle is $\mu_\mathrm{e} \approx 0.6 m_\mathrm{p} \approx 1.0
\times 10^{-24} \mathrm{g}$ and the particle density is $n_\mathrm{e}
= \rho_\mathrm{e} / \mu_\mathrm{e} \approx 2.5 \times 10^{-7}
\mathrm{~cm}^{-3}$. PGC~064440 is in a poor group of galaxies, where
the IGrM particle density is $\sim 10 \times$ as high and the
temperature range is $10^6 \mathrm{~K} < T < 10^7 \mathrm{\,K}$
\citep{sto19}, so the external particle pressure $P_\mathrm{e} = nkT
\approx 10^{-15} \mathrm{~dyne~cm}^{-2}$.
{ Even if $\kappa = 0$, the minimum cocoon pressure is comparable
  with the external pressure.  Higher $\kappa$ or
  non-equipartition magnetic fields would only increase the cocoon
  pressure.}

In the limit $P_\mathrm{e} \ll P_\mathrm{c}$, inserting $P_\mathrm{c}
= P_\mathrm{min}$ into Equation~\ref{eqn:ram} predicts that the cocoon
boundary should be expanding into the surrounding medium with speeds
between $v_\perp \sim 200 (1+\kappa)^{2/7} \mathrm{~km~s}^{-1}$ (IGrM)
and $v_\perp \sim 700 (1+\kappa)^{2/7} \mathrm{~km~s}^{-1}$
(IGM). These expansion speeds are subsonic in the radio cocoons,
allowing the cocoons enough time to reach pressure equilibrium
\citep{beg84} and attain their constant brightness temperatures. The
PKS~2014$-$55 cocoons are $l_\perp \approx 50$~kpc wide, so
{ the expansion time scales $\tau \equiv l_\perp / v_\perp$ of the cocoons are
  $70(1+\kappa)^{-2/7} \la \tau\mathrm{(Myr)} \la 250(1+\kappa)^{-2/7}$.
}

The energy density of the CMB surrounding PKS~2014$-$55 at $z = 0.06$
is $U_\mathrm{CMB} = 5.3 \times 10^{-13} \, \mathrm{erg~cm}^{-3}$. It
is larger than the magnetic energy density in the cocoon even in the
unlikely event \citep{bec05} that $\kappa > m_\mathrm{p} /
m_\mathrm{e} \sim 2000$.  The ratio of IC losses to synchrotron losses
is $U_\mathrm{CMB} / U_B$, so the radiative lifetimes $\sim
100\,\mathrm{Myr}$ of relativistic electrons in the cocoons are
strongly limited by IC scattering off the CMB, not by synchrotron
radiation.

\subsection {The Host Galaxy PGC~064440}\label{subsec:host}

The total stellar mass $M_*$ of PGC~064440
can be estimated photometrically from its $\lambda = 2.17\,\mu$m
absolute magnitude $M_K$ using the empirical relation \citep{cap13}
\begin{equation}
  \log_{10} (M_*/\mathrm{M}_{\sun}) \approx 10.58 - 0.44 (M_K + 23)~.
\end{equation}
The 2MASX apparent magnitude of PGC~064440 is $m_K = 11.64$, the
distance modulus is 32.16, and the $K$ correction is $K
\approx 6.0 \log_{10}(1+z) \approx - 0.15$~mag \citep{koc01}, so $M_K
\approx -25.37$ and the total mass of stars inside the photometric
radius $r \sim 34 \mathrm{\,kpc}$ is $\log_{10} (M_*/\mathrm{M}_{\sun}) \approx
11.6$.

The mid-infrared source WISEA J201801.29$-$\allowbreak553931.5
\citep{wri10} coincides with PGC~064440.
Its mid-infrared colours $(W2-W3) = +3.066$ and $(W1-W2) = +1.256$ are
typical of Seyfert galaxies and far from the colors of elliptical
galaxies whose mid-infrared emission is dominated by stellar
photospheres \citep{jar11}. This mid-infrared evidence for
circumnuclear dust heated by an active galactic nucleus (AGN) is
supported by the presence of a heavily obscured (column density
$\log[N_\mathrm{H} (\mathrm{cm}^{-2})] = 23.51 \pm 0.14$) hard X-ray
source at the centre of PKS~2014$-$55 \citep{pan16} and the absence of
broad optical emission lines. Star formation may also contribute to
the mid-infrared emission from PGC~064440.

PGC~064440 is a Seyfert II galaxy with very strong high-ionization
([\ion{O}{iii}]/H$\beta \sim 13$) emission lines \citep{sim96}.
Many powerful radio galaxies have extended emission-line regions
(EELRs) with radii $\ga 100~\mathrm{kpc}$. \citet{tad89} observed
the [\ion{O}{iii}]$\lambda5007$ line of PGC~064440 with a long slit
in $PA = 192^\circ$ nearly parallel to the continuum major axis and
found emission extending $\sim 11 \mathrm{~kpc}$ on both sides of the
nucleus with the linear velocity field of a constant-density enclosed
mass and maximum rotation velocity $\vert\Delta V^\mathrm{max}\vert
\approx 280 \mathrm{~km~s}^{-1}$ relative to the nucleus, indicating a
total mass $M \sim 2 \times 10^{11}\,\mathrm{M}_{\sun}$ within $ \sim 11
\mathrm{~kpc}$ of the nucleus.

\section{The Hydrodynamical Backflow Model}\label{sec:model}

The extended radio jets of most high-luminosity sources are linear and
inversion symmetric about their host galaxies. \citet{lea84} noted
that opposing axisymmetric backflows could form a fat disc expanding
laterally where they encountered the hot ISM of their host galaxy, but
a misaligned ellipsoidal hot gas halo could break the axial symmetry
and bend the backflows in opposite directions away from the ellipsoid
major axis to produce the secondary arms or ``wings'' of XRGs
(Fig.~\ref{fig:leahy84}).
{ \citet{sar09} found that XRGs lacking FR~II hotspots often contain
  inner doubles, indicating restarted jets along the same axis, and they proposed that
  the wings are asymmetric backflows.
  }

Backflows from extended radio jets
{ can be treated as fluid flows}
because their magnetic fields are strong enough that even
ultrarelativistic protons and electrons have Larmor radii much smaller
than the jet radius. Magnetic lines of force are frozen into the jet
fluid, so velocity shear across the jet tends to align the magnetic
field along the jet, and jet growth can increase the frozen-in magnetic
field strength to near equipartition. Many astrophysical jets are
stable enough to survive bending, as demonstrated by the bent tails of
radio galaxies moving through dense gas in massive galaxy clusters.

\subsection{Backflow Geometry}

\begin{figure}
  \includegraphics[trim={1.7cm 5.0cm 1.5cm 4.5cm}, clip, 
    width=1.\columnwidth]{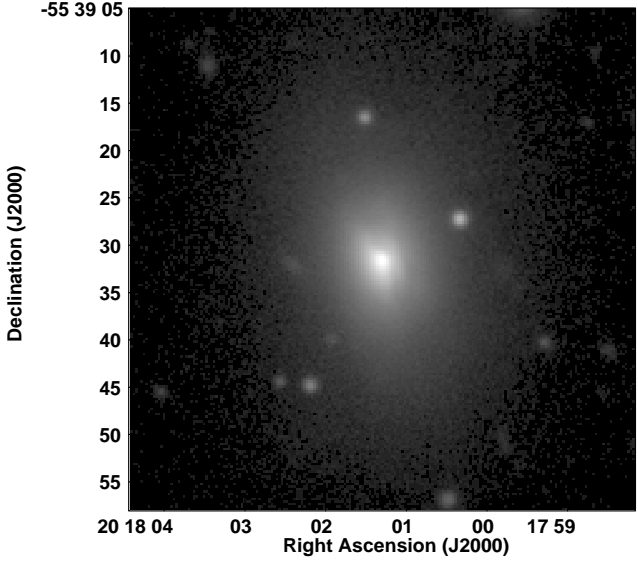}
  \caption{The DES DR1 $r$-band optical image of PGC064440 reveals a faint
    dust lane just northwest of the nucleus.}
  \label{desred}
\end{figure}

\begin{figure}
  \includegraphics[trim={1.7cm 5.0cm 1.5cm 4.5cm}, clip, 
    width=1.\columnwidth]{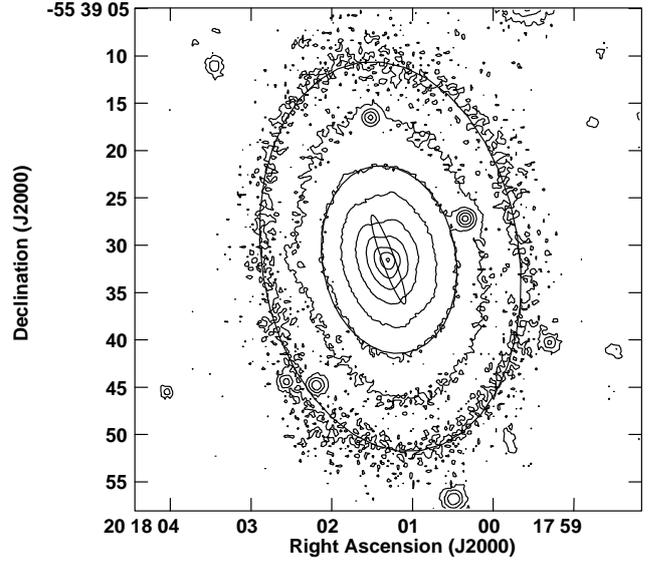}
  \caption{The DES $r$-band brightness contours are shown at $16
    \mathrm{~electrons} \times 1,\, 2,\, 4,\, 8,\, 16,\, 32,\, 64,\,
    128,$ and 256. The $10\arcsec \times 1\farcs3$ ellipse in $PA \approx
    20^\circ$ at the centre represents the dust lane as a tilted
    ring, only the northwestern near side of which is visible in
    absorption. }
  \label{desredcnt}
\end{figure}

A faint dust lane extending $\sim 10\arcsec$ ($\sim 12
\mathrm{\,kpc}$) is just visible in the DES DR1 \citep{abb18} $r$-band
image of PGC~064440 (Fig.~\ref{desred}). Fig.~\ref{desredcnt} is the
corresponding $r$-band brightness contour plot. The narrow inner
ellipse represents the tilted circular ring that overlaps the dust
lane, of which only the near half is visible in absorption. The two
larger ellipses are fits to the first and third brightness contours in
Fig.~\ref{desredcnt}. The parameters of these ellipses and the 2MASS
$K_\mathrm{s}$ ``total'' isophotal ellipse are listed in
Table~\ref{tab:ellfits}. The isophotal ellipticities $\epsilon \equiv
1 - (\phi_\mathrm{m} / \phi_\mathrm{M})$ are in the range $0.28 <
\epsilon < 0.36$. They indicate that PGC~064440 is an oblate ellipsoid
whose equatorial position angle is $PA = 15^\circ \pm 5^\circ$ and
whose projected polar axis is at $PA = 105^\circ \pm 5^\circ$. If the
dust lane is an equatorial ring, the polar axis is inclined $\approx
8^\circ$ from the plane of the sky, with the $PA = 105^\circ$ side
closer to us and the $PA = -75^\circ$ side farther away. Even if PGC
064440 were a thin disc, the isophotal ellipticities independently
imply that its polar axis must be $\la 45^\circ$ from the plane of the
sky.

\begin{table}
\centering
  \caption{Best-fitting ellipses for the dust lane and outer brightness contours
  of PGC~064440.}
  \label{tab:ellfits}
  \begin{tabular}{l c c c}
                              \hline
 & $\phi_\mathrm{M} \times \phi_\mathrm{m}$ & $PA\,$  & $\cos^{-1}(\phi_\mathrm{m}/\phi_\mathrm{M})$\\
   Ellipse     & ($\arcsec ~\times~ \arcsec$) &  (deg) &  (deg) \\
    \hline
 Dust ring & $10 \times 1.3$ & $+20$ & 82\\
 First $r$ contour &  $42 \times 27$ & ~\,$+9$ & 50  \\
 Third $r$ contour &  $20 \times 14$ & $+10$ & 46 \\
 $K_\mathrm{s}$ total &  $57.8 \times 41.6$ & $+15$ & 44 \\
    \hline
  \end{tabular}
\end{table}

\begin{figure}
  \includegraphics[trim={1.5cm 3.0cm 1.5cm 2.0cm}, clip, 
    width=1.\columnwidth]{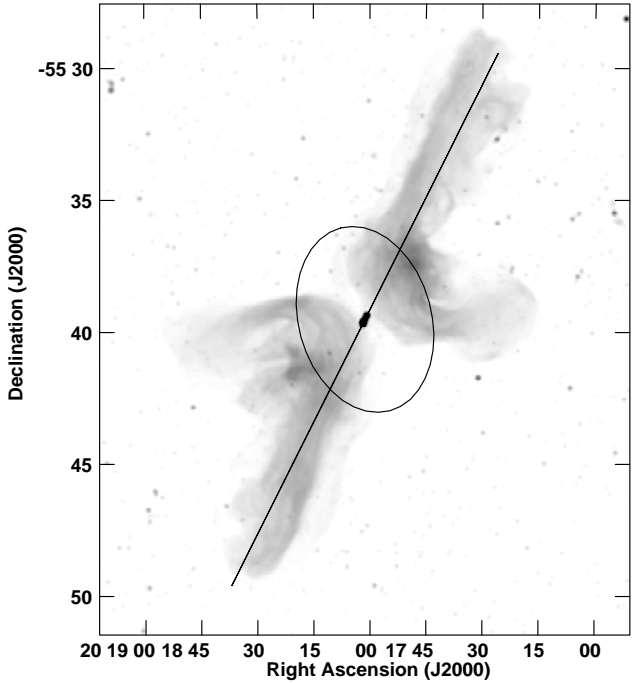}
  \caption{ The straight line segment centred on PGC~064440 is
    1.57~Mpc long at the $PA = 154^\circ$ of the line connecting the
    two inner radio lobes. The ISM of PGC~064440 is represented by an
    ellipse having an $l = 250$~kpc semimajor axis at the same $PA =
    15^\circ$ and ellipticity $\epsilon = 0.3$ as the distribution of
    stellar mass traced by the $K_\mathrm{s}$ isophote
    (Table~\ref{tab:ellfits}). The angle between the jet and the major
    axis of the ellipse is $\Delta = 15^\circ + (180-154)^\circ =
    41^\circ$. }
  \label{fig:geom}
\end{figure}

Fig.~\ref{fig:geom} represents the hypothetical hot halo of PGC~064440 with
an ellipse superimposed on a grey-scale image of PKS~2014$-$55. The
ellipse has the same central position, ellipticity $\epsilon \equiv
(\phi_\mathrm{M} - \phi_\mathrm{m}) / \phi_\mathrm{M} = 0.3$, and
major-axis position angle $PA = 15^\circ$ as the starlight of PGC
064440 traced by its outer $K_\mathrm{s}$ isophote
(Table~\ref{tab:ellfits}). The shape and orientation are justified by
{\it Chandra X-ray Observatory} images \citep{hod10} of hot
($0.3\,\mathrm{keV} < k T < 2 \,\mathrm{keV}$) halo gas surrounding
the host galaxies of XRGs which show that the ellipticities and
position angles of the hot gas follow those of the stellar light
distributions.

The 1.57~Mpc long straight line in Fig.~\ref{fig:geom} is centred
on PGC~064440 and has the same position angle $PA = 154^\circ$ as the
inner triple source. The fact that it closely overlaps the long arms
of PKS~2014$-$55 proves that the radio jet orientation today is within
$\sim 5^\circ$ of its orientation when the jets were launched tens of
millions of years ago
{and when the faint cocoons were formed even earlier.} This is
expected if the radio ``wings'' in $PA \approx 90^\circ$ are produced
by hydrodynamical backflows from stable jets but improbable for models
in which the wings are produced by changing the jet orientation.

{ The hydrodynamical model predicts that the ends of the wings
should contain the ``oldest'' { synchrotron}
electrons. Their very steep continuum spectra (Figure~\ref{fig:SI}) are
consistent with synchrotron and inverse-Compton energy losses.}
{ The cocoons are probably even older.}

The biggest challenge to the hydrodynamical backflow model presented
by PKS~2014$-$55 is the need to deflect both of its very wide
(observed width $\sim 150 \mathrm{\,kpc}$, which we treat as a
cylinder of radius $r \sim 75 \mathrm{~kpc}$) backflows cleanly in
opposite directions without splitting them. This requires that the
hot ISM have both a high ellipticity and a large semimajor axis $l$.
\citet{cap02} found that XRG host galaxies in a small sample all have
$\epsilon > 0.17$, a criterion easily satisfied by PGC~064440, while
only three in their reference sample of 15 3C FR\,II galaxies have
$\epsilon \geq 0.3$, the ellipticity of PGC~064440.

When they encounter the hot halo ISM of the host galaxy, backflows
initially parallel to the main jets are deflected toward the most
negative pressure gradient. The ellipse in Fig.~\ref{fig:geom}
represents an isobaric contour, so backflows bend away from the
directions in which the angle $\Delta$ between the backflow and the
constant-pressure contour of the ellipse is $<90^\circ$.
For any $\Delta$, backflow radius $r$, and ellipticity $\epsilon$,
this implies a minimum galaxy semimajor axis $l_\mathrm{min}$ is needed to
deflect the entire backflow to one side.
In the case of PGC~064440, $\Delta = 41^\circ$ and $\epsilon = 0.3$,
so $x_i \approx 0.854 l$ and $y_i \approx 0.364 l$.
Then Equation~\ref{eqn:minsize} yields the minimum semimajor axis
$l_\mathrm{min}$ that can cleanly deflect the backflow
\begin{equation}
  l_\mathrm{min} = (x_i \sin \Delta - y_i \cos \Delta)^{-1} r
   \approx 3.5 r \approx 250 \mathrm{~kpc}~,
\end{equation}
the size of the ellipse drawn in Fig.~\ref{fig:geom}.
{ This ellipse is just the projection of the galaxy onto the sky,
  which is an ellipsoid of revolution. However, the dust lane
  indicates that the polar axis of PGC~064440 lies only $8^\circ$ from
  the plane of the sky, so the observed ellipse is a good representation of
  the ellipsoid.}

PGC~064440
{ includes} a stellar mass $M_* \approx
10^{11.6} \mathrm{M}_{\sun}$ (Section~\ref{subsec:host}), so the galaxy
stellar-to-halo mass relation (SHMR) predicts that its halo virial
mass should be $M_\mathrm{vir} \ga 10^{13.3} \mathrm{M}_{\sun}$
\citep{wec18}. Such massive galaxies are typically assembled at
redshift $z_\mathrm{a} \sim 0.9$ and have virial radii \citep{shu14}
\begin{equation}
  \biggl( \frac {R_\mathrm{vir}} {206\mathrm{\,kpc}} \biggr) \approx
  \biggl( \frac {M_\mathrm{vir}}{10^{12}\mathrm{M}_{\sun}}\biggr)^{1/3} 
  (1 + z_\mathrm{a})^{-1}~.
\end{equation}
Thus PGC~064440 has $R_\mathrm{vir} \sim 290 $\,kpc, and its halo is
(just) big enough to completely deflect the wide backflows of
PKS~2014$-$55.

\subsection{ Head-tail brightness features in the backflows}

\begin{figure}
  \includegraphics[trim={1.5cm 9.0cm 1.5cm 8.0cm}, clip, 
    width=1.0\columnwidth]{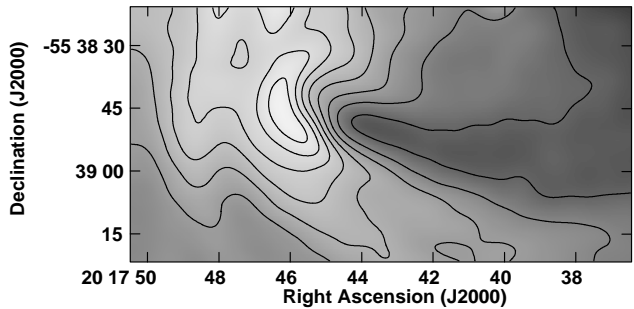}
  \caption{The bright bent ``head'' and long dark ``tail'' in this
    continuum image suggest blockage of the backflow by an obstacle of
    size $\sim 10$~kpc. The contour levels are $200 \,\mu\mathrm{Jy\,beam}^{-1}
  \times 1,\,2,\,3,\dots,8$.}
  \label{fig:bowshock}
\end{figure}

Radio continuum features with bright bent ``heads'' and long dark
``tails'' pointing downstream in the backflows (Fig.~\ref{fig:ipol})
of both secondary lobes at first suggest obstacles blocking the
backflows. The clearest example is in the western backflow
(Fig.~\ref{fig:bowshock}), and there are three weaker examples in the
eastern backflow near J2000 $\alpha =
20^\mathrm{h}\,18^\mathrm{m}\,18\farcs9$, $\delta = -55^\circ\,
39\arcmin\,45\arcsec$, $\alpha =
20^\mathrm{h}\,18^\mathrm{m}\,22\farcs3$, $\delta = -55^\circ\,
39\arcmin\,35\arcsec$, and $\alpha =
20^\mathrm{h}\,18^\mathrm{m}\,22\farcs3$, $\delta = -55^\circ\,
41\arcmin\,04\arcsec$. The apparent width of the dark tails is
slightly larger than the $7\,\farcs4$ resolution of the radio image.
The corresponding linear diameter $\sim$ 10 kpc is comparable with the
size of cold \ion{H}{i} discs in spiral galaxies. However, we detected
no \ion{H}{i} line emission and found no visible galaxies on the DES
images downstream of the heads. Furthermore, 10~kpc is much smaller
than the $2r \sim 150$~kpc diameter of the backflows. The
line-of-sight depth of the backflows from such wide axially symmetric
jets should be $\gg 10$~kpc, so a 10~kpc obstacle could not reduce the
backflow brightness by the observed amount shown in
Fig.~\ref{fig:bowshock}, which is more than a factor of two.

The dark tail in Fig.~\ref{fig:bowshock} cannot be attributed to
free-free absorption in an intervening ionized cloud with kinetic
temperature $T \ga 10^4$~K because an absorber with optical depth
$\tau$ emits with brightness temperature $T_\mathrm{b} = T[1 -
  \exp(-\tau)]$, and the 1.28~GHz brightness temperature $T_\mathrm{b}
\sim 2 \,\mathrm{K}$ of the tail implies $\tau < 10^{-3}$.

Fig.~\ref{fig:geom} shows that the dark tails appear in the
downstream sides of both backflows near the deflecting halo boundary
indicated by the ellipse. It also shows two matching bright regions,
one just inside the apex of each ``boomerang''. These bright regions
are probably not traditional FR\,II hotspots because they are not in
the radio jets and they do not have the usual edge-brightened
morphology. We suggest that the bright regions indicate compression
and turbulence where backflow material is piling up inside the apex.
Turbulence on scales smaller than our $\theta_{1/2} = 7\,\farcs4$ beam
could explain the low observed polarized intensity in the brighter
northwestern region (Fig.~\ref{fig:poln}). We suspect that the
matched sets of dark tails are simply hydrodynamical features
downstream of the bright regions, and don't actually indicate
the presence of external obstructions.

\section{Summary}\label{sec:summary}
The hydrodynamical backflow model for PKS~2014$-$55 is supported by
the following evidence:
\begin{enumerate}
\item{The observed ``double boomerang'' radio morphology is expected
  for backflows from a pair of collinear jets redirected by the
  oblique hot ISM of the host galaxy. Both the magnetic fields and
  the total-intensity ridges follow the continuously bending flow
  lines. Two matching bright regions inside the boomerang apexes suggest
  compression and turbulence where backflow material is piling up. Bright
  heads and dark tails appear between both bright regions and their backflow
  wings. They appear to be features in the flow, not signs of obstruction
  by the ISM of nearby galaxies.}
\item{AGN activity in PKS~2014$-$55 has recently restarted
  (Fig.~\ref{fig:xcore}) with the reborn jets in the same direction
  (Fig.~\ref{fig:geom}) as the main lobe. Thus, the secondary wings
  are very unlikely to be the result of a change in the orientation of
  the spin axis of the SMBH.}
\item{The virial halo of the host galaxy PGC~064440 is large enough
  and has the correct position angle to cleanly deflect backflows from
  the wide main jets in the direction observed.}
\end{enumerate}

The unique combination of high surface-brightness sensitivity ($\sigma
\approx 48 \mathrm{~mK}$), high angular resolution ($\theta_{1/2}
\approx 7\,\farcs4$), and dense $(u,v)$-plane coverage of our new
MeerKAT continuum image makes the very extended radio source
PKS~2014$-$55 the best example of an XRG produced by hydrodynamical
backflows from a jet with fixed orientation. The prototypical XRG
NGC~326 has been cited as evidence for jet reorientation following an
SMBH-SMBH merger \citep{mer02}. However, \citet{har19} reobserved the
dumbbell galaxy NGC~326 with the Low-Frequency Array (LOFAR) at 144~MHz
and found faint, extended radio morphological evidence for
hydrodynamical effects related to an ongoing group or cluster merger.
Although their result does not rule out the spin-flip model for NGC~326,
we endorse their caution not to infer jet reorientation in XRGs
lacking deep and detailed radio images.

The new MeerKAT continuum image also revealed faint ($T_\mathrm{b}
\sim 0.5 \mathrm{\,K}$) low-pressure ($P_\mathrm{min} \sim 10^{-14}
\mathrm {\,dyne\,cm}^{-2}$) cocoons with sharp edges and the nearly
constant brightness characteristic of subsonic (in the cocoons)
expansion into the surrounding intra-group medium probed by the giant
source PKS~2014$-$55. This pressure assumes $\kappa=40$. The
pressure in the cocoons could range from $P_\mathrm{min} \sim 10^{-15}
\mathrm {\,dyne\,cm}^{-2}$ if $\kappa = 0$ to $P_\mathrm{min} \sim
10^{-13} \mathrm {\,dyne\,cm}^{-2}$ if $\kappa=2000$.

\section*{Acknowledgements}
The MeerKAT telescope is operated by the South African Radio Astronomy
Observatory which is a facility of the National Research Foundation,
an agency of the Department of Science and Innovation.
The National Radio Astronomy Observatory is a facility of the National
Science Foundation, operated under a cooperative agreement by Associated
Universities, Inc.
This research has made use of the NASA/IPAC Extragalactic Database
(NED), which is funded by the National Aeronautics and Space
Administration and operated by the California Institute of
Technology. 
This research uses services or data provided by the NOAO Data
Lab. NOAO is operated by the Association of Universities for Research
in Astronomy (AURA), Inc. under a cooperative agreement with the
National Science Foundation.
K. Thorat acknowledges support from the Inter-University Institute for
Data Intensive Astronomy (IDIA).
The financial assistance of the South African Radio Astronomy
Observatory (SARAO) towards this research is hereby acknowledged
(www.ska.ac.za).

{ We thank the anonymous referee for numerous valuable suggestions
to improve our manuscript.}




\bibliographystyle{mnras}
\bibliography{Xgal.bib}



\appendix

\section{Minimum size of the deflecting halo}\label{app:minsize}

In the coordinate system with $x$ parallel to the major
axis and $y$ parallel to the minor axis, the equation of the
deflecting ellipse with semimajor axis $l$ and ellipticity $\epsilon$
is
\begin{equation}\label{eqn:ellipse}
  x^2 + \biggl( \frac {y}{1-\epsilon} \biggr)^2 = l^2
\end{equation}
and tangents to the ellipse have slopes
\begin{equation}\label{eqn:ellslope}
  \frac {\mathrm{d}y}{\mathrm{d}x} = - (1 - \epsilon)^2 \frac {x}{y}~.
\end{equation}
If $\Delta$ is the angle between the backflow and the $x$ axis, the
equation of the outermost backflow of radius $r$ is
\begin{equation}\label{eqn:backflow}
  y = x \tan\Delta - \frac{r}{\cos\Delta}
\end{equation}
and the initial slope of the backflow is $\tan\Delta$. At the point
$(x_i,y_i)$ where it intersects the ellipse at a $90^\circ$ angle, the
slope of the ellipse is $-(\tan\Delta)^{-1}$ and
\begin{equation}
  \frac {y_i}{x_i} = (1 -\epsilon)^2 \tan \Delta~.
\end{equation}
The point $(x_i,y_i)$ must also satisfy Equation~\ref{eqn:ellipse} so
\begin{equation}\label{eqn:minsize}
x_i = l [1 + (1-\epsilon)^2 \tan^2\Delta]^{-1/2}~.
\end{equation}


\bsp	
\label{lastpage}
\end{document}